\documentclass[11pt]{article}

\usepackage{amsmath}
\usepackage{amssymb}
\usepackage{latexsym}
\usepackage{graphics}
\usepackage{psfrag,fancyhdr,epsfig}

\newcommand{\Au}{\mathcal{A}}
\newcommand{\Bu}{\mathcal{B}}
\newcommand{\Cu}{\mathcal{C}}
\newcommand{\Du}{\mathcal{D}}

\newcommand{\Hu}{\mathcal{H}}

\newcommand{\Nu}{\mathcal{N}}
\newcommand{\Ou}{\mathcal{O}}

\newcommand{\Uu}{\mathcal{U}}

\usepackage{color}

\definecolor{orange}{rgb}{0.9,0.2,0}
\definecolor{brown}{rgb}{0.7,0.3,0.2}
\definecolor{fuxia}{rgb}{1,0,1}
\definecolor{skyblue}{rgb}{0,0.1,0.9}
\definecolor{violetred}{rgb}{0.8,0.13,0.56}
\definecolor{deeppink}{rgb}{1.00,0.08,0.5}
\definecolor{pink}{rgb}{1.00,0.75,0.80}
\definecolor{orchid}{rgb}{0.85,0.44,0.84}
\definecolor{lightpink}{rgb}{1.00,0.71,0.76}
\definecolor{bluish}{rgb}{0,0.6,0.8}
\begin{document}

\begin{titlepage}

\begin{centering}

\hfill hep-ph/yymmnnn\\

\vspace{1 in}
{\large{\bf {Bimaximal mixing from lopsided neutrinos.} }}\\
\vspace{1 cm}
{M. Paraskevas
 and K. Tamvakis}\\
\vskip 0.5cm
{$ $\it{Physics Department, University of Ioannina\\
GR451 10 Ioannina, Greece}}

\vskip 0.5cm

\vspace{1.5cm}
{\bf Abstract}\\
\end{centering}
\vspace{.1in}

We consider the problem of neutrino masses and mixing within the general framework of standard (type-I) seesaw models leading to three light neutrinos. Under the assumption of a  hierarchical neutrino mass spectrum  $\lambda^4:\lambda:1$, consistent with present data, we examine possible lopsided patterns for the neutrino Yukawa couplings that can account for the observed mixing angles, including a small but non-vanishing $|U_{e3}|$. An embedding of the above within a general class of $SO(10)$ models is also considered.

 \vfill

\vspace{2cm}
\begin{flushleft}

February 2012
\end{flushleft}
\hrule width 6.7cm \vskip.1mm{\small \small}
 \end{titlepage}

\section{Introduction}
The observation of neutrino oscillations has confirmed, through a number of independent experiments\cite{A}, not only the existence of at least two massive neutrinos  but also a mismatch between flavour and mass eigenstates in the lepton sector. This fact by itself provides a first piece of compelling evidence for physics beyond the Standard Model(SM).
Present data\cite{RPP}, although not yet conclusive on the full neutrino mass spectrum, favour a three generation scenario for the light neutrinos at an overall mass scale $M\sim10^{-1}eV$. Based on the squared mass differences $\delta m_{12}^2\approx 7.65\times 10^{-5} eV^2 ,\,\delta m_{23}^2\approx 2.4\times 10^{-3} eV^2$   which are currently determined by experiment, three distinct possibilities  may  rise, depending on the explicit hierarchical spectrum one assumes. In the {\textit{Normal Hierarchy}} (NH) case, for  $m_1\ll(\delta m_{12}^2)^{1/2}$, one obtains $m_1\ll m_2\approx (\delta m_{12}^2)^{1/2},\, m_2\ll m_3\approx (\delta m_{23}^2)^{1/2}$, while in the {\textit{Inverse Hierarchy}} (IH) case, for  $m_3\ll(\delta m_{23}^2)^{1/2}$, one obtains $m_3\ll m_1\approx m_2\approx (\delta m_{23}^2)^{1/2}$. An almost degenerate spectrum is also possible for $m_1\approx m_2 \approx m_3\gtrsim 10^{-1}eV\,(\lesssim 1\,eV)$. In the last two cases of partial or complete degeneracy the mass spectrum exhibits a tuned form and the relation  $\delta m_{12}^2 \ll\delta m_{23}^2$ seems accidental. In contrast, such a relation seems more natural in the NH scenario and thus adds to its attractiveness. Nevertheless, this has to be reconciled with the large mixing angles observed in neutrino oscillations.

Lepton mixing\cite{Dorsner} in the framework of the SM can arise if one considers general mass matrices for the neutrinos. The mass matrix of the light neutrinos and that of the charged leptons diagonalize simultaneously by distinct unitary and biunitary transformations leaving a physical trace on the charged weak current. In a way analogous to quark mixing in the CKM matrix, lepton mixing can be described by $U_{PMNS}\equiv U_e^\dag U_\nu $ where $U_e,U_\nu$ are the unitary transformations of the left-handed charged leptons and the light neutrinos respectively. Not all parameters in this expression are observable. In general, the physical parameters of $U_{PMNS}$  can be expressed in terms of three real angles and 1 or 3 CP-violating phases in case of Dirac or Majorana neutrinos respectively\cite{RPP}\cite{Valle}\cite{Fritzsch}. At present, neutrino oscillation experiments favour a nearly bimaximal pattern for the 2 real mixing angles and a small but non-vanishing value for the third ($\lesssim 12^{o}$).

 Focusing on the issue of reconciling a normal mass hierarchy with large mixing, one finds that for a symmetric hierarchical neutrino mass matrix, the corresponding unitary transformation $U_\nu$ is likely to contribute small angles to $U_{PMNS}$ and in fact of comparable or smaller magnitude than the relevant mass ratios. Then, a pragmatic approach to solve the problem ``large mixing-large hierarchy" would be to assume large asymmetric (lopsided) elements in the mass matrix of the charged leptons\cite{Barr,Barr2}. In this way the observed bimaximal mixing originates from $U_e^\dag$ and the hierarchical mass matrix of the light neutrinos contributes  small or negligible corrections through $U_\nu$.
Note however that the above initial argument is not general. Symmetric matrices may accommodate a large hierarchical spectrum and at the same time contribute large mixing angles without fine-tuning.  The general idea is that the symmetric light neutrino matrix may have an underlying lopsided substructure. The seesaw mechanism, besides explaining the smallness of the overall light neutrino mass scale, seems to provide us with a suitable framework for this to happen in a natural way. In this approach, based on the attractive properties of the lopsided models, the observed bimaximal mixing can be partially or completely accounted for by the neutrino sector\cite{Sato}.

In what follows we examine analytically a number of lopsided ansatze for the lepton sector that can potentially fit current low energy data. Large lepton mixing is raised from both the charged leptons and the neutrinos or through the neutrino sector exclusively, as in the case of a particularly simple ansatz, which is investigated thoroughly. We explore the possibility of embedding this pattern within a class of $SO(10)$ models with realistic fermion masses and mixings. In section 2 we illustrate the general features of lopsided models and their relation to large mixing. In section 3 we discuss briefly the standard Type-I seesaw framework and present our conventions. In section 4 we consider and study a number of lopsided patterns that lead to the observed lepton mixing. In section 5 we concentrate on a particularly simple ansatz that leads to lepton mixing exclusively through the neutrino sector. In section 6 we consider the embedding of the above in a class of $SO(10)$ models and, finally, in section 7 we state our conclusions.

\section{Lopsided models and large mixing.}
In order to illustrate some of the main features of lopsided models we may consider a general matrix $Y_{ij}\,=\,C_{ij}\,\epsilon_j$
\begin{equation}Y\,=\,\left(\begin{array}{ccc}
C_{11}\epsilon_1\,&\,C_{12}\epsilon_2\,&\,C_{13}\\
\,&\,&\\
C_{21}\epsilon_1\,&\,C_{22}\epsilon_2\,&\,C_{23}\\
\,&\,&\\
C_{31}\epsilon_1\,&\,C_{32}\epsilon_2\,&\,C_{33}
\end{array}\right)\end{equation}
 with a hierarchy of the form
 \begin{equation}\epsilon_1\ll\epsilon_2\ll\epsilon_3\,\equiv\,1\end{equation}
  and random $\mathcal{O}(1)$ coefficients $C_{ij}$, taken real for simplicity. Diagonalization proceeds as usual with the biorthogonal transformation
\begin{equation}Y_\Du\,=\,U_1^\bot\, Y\, U_2\,=\,U_2^\bot \,Y^\bot\, U_1\end{equation}
 but with $U_1$ including large $\mathcal{O}(1)$ rotation angles, while those of $U_2$ are small. Depending on the explicit  hierarchical form of the  matrix the largest rotation angle inside $U_2$ may be $\mathcal{O}(\epsilon_2)$ or $\mathcal{O}(\epsilon_1 /\epsilon_2)$. If $U_1$ participates in the PMNS matrix and $U_2$ in the CKM matrix, as is the case in standard $SU(5)$, where $Y\equiv Y^{(d)}=(Y^{(e)})^\bot$, large angles will be attributed to the former and small to the latter.

Another attractive aspect of lopsided matrices is that they can produce symmetric matrices that can both accommodate a hierarchical spectrum and large mixing angles in a natural way. Since
\begin{equation}
Y_\Du^2\,=\,U_1^\bot\, Y\, Y^\bot \,U_1\,=\,U_2^\bot\, Y^\bot\, Y \,U_2\label{Yd2}
\end{equation}
both symmetric matrices $YY^\bot$ and $Y^\bot Y$ share the same eigenvalues\cite{Valle2}. In fact, it is much easier to extract the mass eigenvalues from $Y^\bot Y$ which diagonalizes with small angles due to its hierarchical form. On the other hand $Y Y^\bot$ can be reexpressed as
\begin{equation}
Y Y^\bot = \Au + \epsilon_2^2 \Bu +\epsilon_1^2 \Cu\,,{\label{X}}
\end{equation}
where $\Au,\,\Bu,\,\Cu$ are symmetric rank-1 matrices. First we diagonalize $\Au$ with $U_\Au=U_{12}U_{23}$ where\footnote{ We use the  notation $\tan_{ij}\equiv\tan{\theta_{ij}} $ for  trigonometric functions where subscripts indicate rotations in the respective planes of family space.} $$\tan_{12}=C_{13}/C_{23},\,\,\,\tan_{23}\,=\,\frac{(C_{13}^2+C_{23}^2)^{1/2}}{C_{33}}$$
 and, thus,
 \begin{equation}
 U_\Au ^\bot \,Y \,Y^\bot\, U_\Au\, = \,\Au_\Du \,+\, \epsilon_2^2 \Bu'\, +\,\epsilon_1^2\, \Cu',\,\,\,\,
 \Au_\Du\,=\,\left(\begin{array}{ccc}
0 & 0 & 0 \\
0 & 0 & 0 \\
0 & 0 & \sum_kC_{k3}^2
\end{array} \right)\,.
\end{equation}
 We should note that there is no reason for the rotated $\Bu'$ or $\Cu'$ to be diagonal. In fact, such a tuned case would correspond to proportional coefficients inside $Y$ and thus imply a rank-2 or even a rank-1 form.
Next, we rotate with $U_{\Bu'}\,=\,U'_{12}$, where now
$\tan'_{12}\,=\,C_{12}'/C_{22}'$, and obtain\footnote{Primed coefficients correspond to the elements of the rotated matrices. The explicit expressions are $$
C_{12}'\,=\,C_{12}\,\cos_{12}\,-C_{22}\sin_{12},\,\,\,\,\,\,\,\,\,\,C_{22}'\,=\,\left(C_{12}\sin_{12}\,+\,C_{22}\cos_{12}\right)\cos_{23}\,-C_{32}\sin_{23}\,.$$ }
\begin{equation}U_{\Bu'}^\bot \,U_\Au ^\bot \,Y\, Y^\bot\, U_\Au\, U_{\Bu'}\,=\,\Au_\Du \,+\,\epsilon_2^2 \Bu''\,+\,\epsilon_1^2 \Cu'',\end{equation}
\begin{equation}\Bu ''\,=\,\left(\begin{array}{ccc}
0 & 0 & 0 \\
\,&\,&\\
0 & {C_{12}'}^2\,+\,{C_{22}'}^2 & {C_{32}'}({C_{12}'}^2\,+\,{C_{22}'}^2)^{1/2} \\
\,&\,&\,\\
\,0\, &\,{C_{32}'}({C_{12}'}^2\,+\,{C_{22}'}^2)^{1/2} \,& {C_{32}'}^2
\end{array} \right)\,.
\end{equation}
The full rotation matrix can be approximated by $U_1\approx U_\Au U_{\Bu'}=U_{12}U_{23}U'_{12}$ at dominant level which results to a form
\begin{equation}
U_1^\bot YY^\bot U_1\sim\left(\begin{array}{ccc}
\epsilon_1^2 & \epsilon_1^2 & \epsilon_1^2 \\
\,&\,&\,\\
\epsilon_1^2 & \epsilon_2^2 & \epsilon_2^2 \\
\,&\,&\,\\
\epsilon_1^2 & \epsilon_2^2 & 1
\end{array} \right)  {\label{Z}}
\end{equation}
Diagonalization is then completed with subdominant rotations of $\Ou (\epsilon_2^2)$ or $\Ou(\epsilon_1^2/\epsilon_2^2)$.

In the above analysis we considered real coefficients $C_{ij}$ for the  elements of the Yukawa matrices allowing for an analytic treatment of the rotation matrices and eigenvalues. In the general case of complex parameters equation (\ref{Yd2}) is no longer valid but the main properties of lopsided matrices still hold. For example,  $Y\,Y^\bot$ and $Y^\bot\, Y$, which now have different eigenvalues, are diagonalized by $U'_1$ and $U'_2$ respectively. These are different from the  $U_1,\,U_2$  that diagonalize $Y$ directly through $Y_\Du\, =\,U_1^\dag\, Y \,U_2$. Both symmetric matrices though, in general, share a similar hierarchical spectrum and equations ({\ref{X}}-{\ref{Z}}) still hold for analogous complex rotations.

\section{Mass scales and Seesaw.}
Choosing as a general framework the two-doublets SM, electroweak symmetry breaking is realized through a non-vanishing vev for $\Hu_u,\Hu_d$ in the direction of their neutral component. Thus, we obtain the mass terms for the charged fermions
\begin{equation}
v_d Y^{(d)}_{ij}\,d^c_i\, d_{j}\,+\,v_u Y^{(u)}_{ij}\, u^c_i\, u_{j}\,+\,v_d Y^{(e)}_{ij}\,e^c_i\, e_{j}
\end{equation}
and for the neutrinos
\begin{equation}
v_u Y^{(N)}_{ij}\,\nu_i\,\Nu^c_j \,+\,\frac{1}{2}M_R Y^{(R)}_{ij}\,\Nu^c_i\,\Nu^c_j\,.
\end{equation}
For a right handed neutrino mass scale in the neighborhood of  $M_R\sim 10^{14}\, GeV$, a standard seesaw mechanism can be realized leading to the  effective light neutrino mass
\begin{equation}
M_\nu \,\approx\, -\frac{v_u^2}{M_R}\, Y^{(N)} \,{Y^{(R)}}^{-1}\, {Y^{(N)}}^\bot\, \label{Mn}\,.
\end{equation}
Then, the resulting overall mass scale ${(v_u^2/M_R)}$ comes out roughly as $\sim 10^{-1}\, eV$, in agreement with present data. Of course, for the above formula to be valid, $v_u Y^{(N)}_\Du \,\ll\, M_R Y^{(R)}_\Du$ should in general hold for the eigenvalues. If this is not the case, then, heavy $\Ou (M_W)$ Dirac-like masses would be produced reducing the number of light neutrinos.
Under these considerations and neglecting the overall mass scale, eqn.(\ref{Mn}) can be reexpressed in the more convenient form as
\begin{equation}
Y^{(\nu)} \approx Y Y^\bot\,,
\end{equation}
where
\begin{equation} Y\,\equiv\,{Y^{(N)}}\,\left(Y^{(R)}_\Du\right)^{-1/2}\,.\label{Y}
\end{equation}
 This allows us to manipulate neutrino masses and mixings as in eqn.({\ref{X}})-({\ref{Z}}). All Yukawa matrices for fermions are expressed in a basis where the right handed neutrino mass matrix is diagonal with real and positive entries. The definition in eqn.(\ref{Y}) is then straightforward.

A lopsided structure, along with the desired hierarchy, may arise in various ways. For example, if $Y^{(R)}_\Du$ is a diagonal matrix (possibly with a suitable hierarchy), $Y^{(N)}$ can be responsible for the lopsided form of $Y$ and an associated hierarchy, a possibility well motivated by GUT considerations. Alternatively, if one assumes a generic $Y^{(N)}$ with $\Ou (1)$ matrix elements and a hierarchical $Y^{(R)}_\Du$, an analogous lopsided $Y$ can be obtained but in this case a lower bound for the mass of the lightest neutrino is also inherited\footnote{A typical hierarchy $\lambda^4:\lambda:1$ for the light neutrinos, parametrized by $\lambda\equiv(\delta m_{12}^2/\delta m_{23}^2)^{1/2}\,\approx\,0.18$, would in general require an inverse hierarchy $\lambda^{-4}:\lambda^{-1}:1$ for the right handed neutrinos, implying a mass eigenvalue $M_R\lambda^{-4}$ close to the physical cutoff of the theory whether this is the GUT, String or Planck scale. }. In what follows, we will be interested in the explicit form of $Y$ with the remark that the examined patterns can be obtained from the more fundamental matrices $Y^{(R)},\,Y^{(N)}.$

\section{Lopsided Lepton Patterns.}

Next we proceed with the examination of possible lopsided patterns for the matrix $Y$ defined in eqn.({\ref{Y}}) that can contribute large  mixing angles to $U_{PMNS}$ through the neutrino sector. Three working examples are the following $Y_1,Y_2,Y_3$
$$
\left(\begin{array}{ccc}
C_{11}\lambda^2\,&\,C_{12}\lambda^{1/2}\,&\,C_{13}\\
\,&\,&\\
C_{21}\lambda^2\,&\,C_{22}\lambda^{1/2}\,&\,C_{23}\\
\,&\,&\\
C_{31}\lambda^2\,&\,C_{32}\lambda^{1/2}\,&\,C_{33}
\end{array}\right),\,
\left(\begin{array}{ccc}
C_{11}\lambda^2\,&\,C_{12}\lambda^{1/2}&\dots\\
\,&\,&\\
C_{21}\lambda^2\,&\,C_{22}\lambda^{1/2}&\dots\\
\,&\,&\\
C_{31}\lambda^2\,&\,C_{32}\lambda^{1/2}&C_{33}
\end{array}\right),\,
\left(\begin{array}{ccc}
C_{11}\lambda^{2}\,&\,C_{12}\lambda^{1/2}\,&\dots\\
\,&\,&\\
C_{21}\lambda^{2}\,&\,C_{22}\lambda^{1/2}\,&\dots\\
\,&\,&\\
\dots&\dots&C_{33}
\end{array}\right)\,$$
where the dots signify entries smaller than the ones explicitly shown which we can safely neglect, i.e. $\dots\,<<\,O(\lambda^2)$. One should not be alarmed by the half-integer powers of the bookkeeping small parameter $\lambda$, since these matrices correspond, through the see-saw formula, to couplings with integer powers of $\lambda$, as could be expected to arise in various flavour-symmetry breaking schemes. All $Y_i$'s correspond to a typical spectrum  $\lambda^4 :\lambda :1$ in the NH case of the neutrinos, although they can be easily modified to accommodate a smaller value for the mass of the lightest neutrino.

The associated charged lepton matrices  $Y^{(e)}_1,Y^{(e)}_2,Y^{(e)}_3$ are
$$
\left(\begin{array}{ccc}
\tilde{C}_{11}\kappa^3\,&\,\tilde{C}_{12}\kappa^3\,&\,\dots\\
\,&\,&\,\\
C_{13}\kappa\,&\,C_{23}
\kappa\,&\,\dots\\
\,&\,&\,\\
\dots\,&\,\dots\,&\,\tilde{C}_{33}
\end{array}\right),\left(\begin{array}{ccc}
\tilde{C}_{11}\kappa^3\,&\,\dots\,&\,\dots\\
\,&\,&\,\\
\dots\,&\,\tilde{C}_{22}\kappa\,&\,\tilde{C}_{23}\kappa\\
\,&\,&\,\\
\dots\,&\,\tilde{C}_{32}\,&\,\tilde{C}_{33}
\end{array}\right),\,
\left(\begin{array}{ccc}
\tilde{C}_{11}\kappa^3\,&\,\dots\,&\,\dots\\
\,&\,&\,\\
\tilde{\kappa}\,&\,\tilde{C}_{22}\kappa\,&\,\tilde{C}_{23}\kappa\\
\,&\,&\,\\
\dots\,&\,\tilde{C}_{32}\,&\,\tilde{C}_{33}
\end{array}\right),$$
all corresponding  to the mass hierarchy $\kappa^3 :\kappa :1$ parametrized by the  small parameter $\kappa = m_\mu /m_\tau$ in a manner consistent with current low energy data.

$Y_1$ has been previously used in Section 2 as an example where an arbitrary hierarchy $\epsilon_1^2:\epsilon_2^2 : 1$ was assigned to $YY^\bot$. By substituting $\epsilon_1^2,\,\epsilon_2^2$ with $ \lambda^4,\lambda $ respectively we obtain the desired neutrino hierarchy and the unitary transformation $U_\nu\approx U_{12}U_{23}U'_{12}U'_{23}$. Among these unitary matrices only $U'_{23}$ is the subdominant rotation of $\Ou(\lambda)$ needed to complete diagonalization (up to negligible corrections) and all other are $\Ou(1)$. The large mixing angles are explicitly given by the same expressions as before, namely
$$
\tan_{12}\,=\frac{{C}_{13}}{{C}_{23}},\,\, \,\tan_{23}\,=\frac{(C_{13}^2+C_{23}^2)^{1/2}}{C_{33}},\,\,\,\tan'_{12}\,=\frac{({C}_{12}\cos_{12}-{C}_{22}\sin_{12})}{({C}_{12}\sin_{12}+{C}_{22}\cos_{12})\cos_{23}-{C}_{32}\sin_{23}}\,.
$$
 If we neglect the contribution from the charged lepton sector, a direct comparison with the standard parametrization $U_{PMNS}=\Uu_{23}\Uu_{13}\Uu_{12}$ results in three $\Ou(1)$  angles and therefore a trimaximal scheme in disagreement with present observations. Then, in order to fit the mixing angles, perhaps the easiest way is to assume a large contribution from the charged leptons along with a certain amount of fine tuning through the relation $U_e=U_{12}$ with $\tan_{12}\approx {C}_{13}/{C}_{23}$. In this sense $Y^{(e)}_1$ has what is required to obtain  $U_{PMNS}\approx U_{23} U'_{12}U'_{23}$  that fits better the observed mixing pattern.

Using again the formalism developed in eqn.({\ref{X}}-{\ref{Z}}) for $Y_2$ we obtain the unitary transformation $U_\nu\approx U_{12} U'_{23}$, where now the neutrino mixing angles are given by the expressions $$\tan_{12}=\frac{{C}_{12}}{{C}_{22}},\,\,\,\,\,\,\,\,\tan'_{23}\approx \lambda \frac{{C}_{32}}{C_{33}^2}({C}_{12}^2+{C}_{22}^2)^{1/2}\,.$$
 Since only one large angle is obtained in this way, the contribution of the charged leptons is again required but with the apparent advantage that no fine-tuning has to be imposed. Then, $Y^{(e)}_2$ is diagonalized by a rotation of the left-handed fields $U_e=U_{23}$ with an $\Ou(1)$ mixing angle given by $\tan_{23}\approx\tilde{C}_{32}/\tilde{C}_{33}$. The resulting unitary transformation describing lepton mixing will then be $U_{PMNS}\approx U_{23}^\dag U_{12} U'_{23}$, which can easily fit lepton mixing data.

There is a third pattern that can be seen as a variation of the previous one with the difference that now the observed small angle of the lepton mixing originates from the charged lepton sector.  Assuming $Y_3$ for the neutrinos, we obtain $U_\nu=U_{12}$ with $\tan_{12}\approx C_{12}/C_{22}$. On the other hand from $Y^{(e)}_3$, with the additional choice for the new scale $\tilde\kappa\sim \lambda \kappa$ we obtain $U_e\approx U_{23} U'_{12}$ up to $\Ou(\kappa^2)$ corrections with $\tan_{23}=\tilde{C}_{32}/\tilde{C}_{33},\,\tan'_{12}=\tilde{\kappa}/[(\tilde{C}_{22}\cos_{23}-\tilde{C}_{23}\sin_{23}) \kappa]\sim\lambda$. Consequently, lepton mixing is now described by $U_{PMNS}\approx U_{12}^{'\dag} U_{23}^\dag U_{12}$ an expression also consistent with present data.

\section{A Lopsided Neutrino Pattern.}
There is an interesting and attractive possibility that the lepton mixing pattern observed in nature originates solely from the neutrino sector. In this section we shall explore this possibility in the general case of complex $\Ou (1)$  coefficients {$C_{ij}=|C_{ij}|e^{i\phi_{ij}}$}. The related unitary transformations can be parametrized in terms of a real angle and a complex phase. For example, a unitary complex rotation in the $\{12\}$ plane can be described by \footnote{Lepton  mixing can be described by various equivalent parametrizations\cite{Fritzsch}. Nevertheless, the symmetrical parametrization  $\Uu_{23}(\hat{\theta}_{23};\hat{\delta}_{23})\Uu_{13}(\hat{\theta}_{13};\hat{\delta}_{13})\Uu_{12}(\hat{\theta}_{12};\hat{\delta}_{12})$ in the presence of $CP$-violating phases seems more attractive  for  model building purposes\cite{Valle}.}
\begin{equation}U_{12}=\left(\begin{array}{ccc}
\cos_{12} & \sin_{12}e^{-i\delta_{12}} & 0 \\
\,&\,&\\
-\sin_{12}e^{i\delta_{12}} & \cos_{12} & 0 \\
\,&\,&\\
0 & 0 & 1
\end{array}\right)\,.\end{equation}
Let us consider
\begin{equation}
Y\,=\,\left(\begin{array}{ccc}
C_{11}\lambda^2\, &\,C_{12}\lambda^{1/2} \, & \,\dots\\
\,&\,&\,\\
C_{21}\lambda^2\, &\,C_{22} \lambda^{1/2} \, & \,C_{23} \\
\,&\,&\,\\
C_{31}\lambda^2\, &\,C_{32} \lambda^{1/2}\,  & \,C_{33}
\end{array}\right)\label{Yn}
\end{equation}
for the matrix of the neutrinos defined in ({\ref{Y}}), which, as before, corresponds to a typical hierarchy $\lambda^4:\lambda:1$ of the NH case. Furthermore, we assume a negligible contribution to lepton mixing from the charged lepton sector, an assumption  motivated by the large mass hierarchy of the charged leptons. This covers a large variety of distinct realistic patterns for the $Y^{(e)}$'s. In this sense we can have to a good approximation $U_e\approx I$ and, as a result, the useful property that diagonal phase matrices commute with $U_e$.

Diagonalization of the neutrino matrix proceeds as usual through the formalism developed in ({\ref{X}})-({\ref{Z}}). Note that by a field redefinition we can absorb the complex phases of $C_{12},\,C_{23},\,C_{33}$. Diagonalization then  begins with the unitary transformation $U_{23}U_{12}$ describing two successive rotations. The corresponding large rotation angles and the complex phases  are given by
\begin{equation}
\tan_{23}=C_{23}/C_{33},\,\, \delta_{23}=0\,\,,\,\,\tan_{12}=C_{12}/|C_{22}'|\,,\delta_{12}=-\phi_{22}'
\end{equation}
with the complex primed coefficients  
\begin{equation}
C_{22}'\,=\, C_{22}\cos_{23}\, -C_{32}\sin_{23},\,\,\,\,C_{32}'\,=\, C_{22}\sin_{23}\,  + \,C_{32}\cos_{23}\,.
\end{equation}
Thus, the neutrino mass matrix $YY^\bot$ is brought into the hierarchical form
$$
\left(\begin{array}{ccc}
{C'}_{11}^2\lambda^4 \,&\, (\dots)\lambda^4 \, & \,(\dots)\lambda^4 \\
\,&\,&\,\\
(\dots)\lambda^4\, & \,e^{2i\phi_{22}'}\left(C_{12}^2+|C_{22}'|^2\right)\,\lambda\, & \, e^{i(\phi_{22}'+\phi_{32}')}|C_{32}'|\left(C_{12}^2+|C_{22}'|^2\right)^{1/2}\,\lambda \\
\,&\,&\,\\
(\dots)\lambda^4\, &\,\, e^{i(\phi_{22}'+\phi_{32}')}|C_{32}'|\left(C_{12}^2+|C_{22}'|^2\right)^{1/2}\,\lambda\, &\,C_{23}^2\,+\,C_{33}^2\,+\,\lambda {C_{32}'}^2
\end{array}\right)\,.
$$
The coefficients  denoted by dots and multiplying the $\lambda^4$ elements are irrelevant since only $C'_{11}$ is in practice associated  with the lightest neutrino mass and a contribution to the CP-violating phases\footnote{$C_{11}'$ is given explicitly by $C_{11}'=C_{11}\cos_{12}-(C_{21}\cos_{23}-C_{31}\sin_{23})\sin_{12}e^{-i\phi_{22}'}$.}. A subsequent small complex rotation $U'_{23}$, with
$$\tan'_{23}=\lambda |C_{32}'|\frac{(C_{12}^2+|C_{22}'|^2)^{1/2}}{(C_{23}^2+C_{33}^2)}\,+\,\Ou(\lambda^2),\,\,\,\,\,\,\,\,\,\delta'_{23}=\phi_{22}'+\phi_{32}'+\Ou(\lambda)\,,$$ along with negligible $\Ou (\lambda^3)$ rotations, will finally bring the neutrino matrix to the diagonal form

\begin{equation}
Y^{(\nu)}_\Du \,\approx\,
  \left(\,
\begin{array}{ccc}
{C_{11}'}^2\lambda^4 \,&\,0\,&\,0\\
\,&\,&\,\\
 \,0\,&\,e^{2i\phi_{22}'}\left(C_{12}^2+|C_{22}'|^2\right)\,\lambda\,&\,0\\
 \,&\,&\,\\
 \,0\,&\,0\,&\, C_{23}^2\,+\,C_{33}^2\,+\,\lambda {C_{32}'}^2\,\\
 \,&\,&\,
 \end{array}\right)
\,.
\end{equation}
Summarizing, lepton mixing in this model is described by the unitary transformation
\begin{equation}
U_{23}(\theta_{23},\,0)\,U_{12}(\theta_{12},-\phi_{22}')\,U'_{23}(\theta'_{23},\,\phi_{22}'+\phi_{32}')\,\cdot\,{\cal{ P}},\,\label{U}
\end{equation}
with
\begin{equation}
{\cal{P}}\,\approx\,diag\left(e^{-i\phi_{11}'},\,e^{-i\phi_{22}'},\,1\right)\,.
\end{equation}
${\cal{P}}$ guarantees the real positive mass eigenvalues. We already notice the predictive power of this pattern. Starting from a general complex
matrix $Y$ for the neutrinos, with 8 complex parameters, and assuming a lopsided structure, consistent with a typical hierarchical spectrum, we obtained an one to one fit between $C_{23},\,C_{33},\,C_{12},|C_{22}|,|C_{32}|,\phi_{22}, \phi_{32}$ and the two heavier neutrino masses, the three rotation angles and the two (out of three) CP-violating phases. Furthermore, the two rotation angles are predicted $\Ou(1)$, while the third is $\Ou(\lambda)$  as a consequence of the neutrino mass hierarchy $\lambda: 1$.

In order to exhibit the explicit relations of observables, we first note that the expression ({\ref{U}}) is unique up to a left-multiplication by an arbitrary diagonal phase matrix. By a field redefinition of the left-handed charged leptons, having assumed that $U_e\approx I$, we  obtain
\begin{equation}
U_{PMNS}\,\approx \,{\cal{P}}^{-1}\,\large{U}_{23}(\theta_{23};0)\,U_{12}(\theta_{12};-\phi_{22}')\,U'_{23}(\theta'_{23};\,\phi_{22}'+\phi_{32}')\,{\cal{P}}\,.
\end{equation}
A direct comparison with the symmetrical parametrization of the physical quantities results in the following relations
\begin{equation}
\tan{\theta}_{sol}\,\equiv\,{\tan}\hat{_{12}}\,\approx\,\tan_{12}\end{equation}
\begin{equation}\tan{\theta}_{atm}\,\equiv\,\tan\hat{_{23}}\,\approx\, |\tan_{23}+ e^{-i(\phi_{22}'+\phi_{32}')}  \cos_ {12}\,\tan'_{23}|\end{equation}
\begin{equation}|U_{e3}|\,\equiv\,\sin\hat{_{13}}\,\approx\,\sin_{12}\,\sin'_{23}\end{equation}
\begin{equation}\hat{\delta}_{12}\,\approx\,-\phi_{11}',\,\,\,\hat{\delta}_{13}\,\approx\,\phi_{32}'\,-\phi_{11}',\,\,\,\,\,\hat{\delta}_{23}\,\approx\, -\phi_{22}'\,.\label{delta}
\end{equation}
where the relations for the phases hold up to $\Ou{(\lambda)}$ corrections.
The Dirac CP-phase of the standard parametrization responsible for CP-violation in neutrino oscillations is identified as $$\delta_D ^{lep}\,\equiv\,\hat{\delta}_{13}\,-\hat{\delta}_{12}\,-\hat{\delta}_{23}
\,\approx\,\phi_{22}'\,+\,\phi_{32}'\,.$$
 Our initial choice of same order parametrization coefficients, so that $C_{ij}\,\sim\,\Ou(1)$, is well justified by fitting the current experimental data from neutrino oscillation phenomena. Nevertheless, the Dirac CP-phase is required for a more accurate fit between the three observed mixing angles, the two heavier neutrino masses, and the subset of the parameters $\left\{C_{23},\,C_{33},\,C_{12},|C_{22}'|,|C_{32}'|\right\}$. A more conclusive test for this model, including the complex phases $\phi_{22},\,\phi_{32}$, would further require the measurement of any existing physical Majorana phases.
Even at this stage however, taking at face value $\sin_{23}'\,\approx\,\lambda$, we arrive at the interesting estimate
\begin{equation}
|U_{e3}|\,\approx\,\lambda\sin{\theta}_{sol}\,\approx\,\sin{5.9^{o}}\,.
\end{equation}
Concluding our discussion, we note some of the general characteristics and perspectives of this pattern. If any or all of the $C_{i1}$'s in (\ref{Yn}) are substituted by texture zeros (or smaller entries) the same relations are obtained up to a different complex phase contribution $\phi'_{11}$ and a different corresponding light neutrino spectrum of the form $(<\lambda^4):\lambda: 1$, something still consistent with observations. If on the other hand, either $C_{22}$ or $C_{32}$ (but not both) are replaced with texture zeros, two additional predictive relations are obtained. By taking $C_{32}$ zero we obtain a straightforward relation for the complex phases in (\ref{delta}) since $\phi'_{22}=\phi'_{32}=\phi_{22}$  and the relation for the mixing angles
\begin{equation}
\tan{\theta_{atm}}\approx\frac{|U_{e3}|}{\tan{\theta_{sol}}}
\left|\frac{(m_{\nu_3}-m_{\nu_2}\cos\delta_D^{lep})}{m_{\nu_2}\cos^2{\theta_{sol}}}+\cos\delta_D^{lep}\right|.\label{Ue3}
\end{equation}
Using current best-fit values for $\tan{\theta_{atm}},\tan{\theta_{sol}},m_{\nu_2}/m_{\nu_3}$, the small angle $\hat{\theta}_{13}$ is predicted in the ($4^o$-$6^o$) region.
For a vanishing $C_{22}$ an analogous relation can be obtained.

\section{Embedding in GUTs.}
In the previous section we showed how a lopsided structure in the neutrino sector may lead to the observed lepton mixing angles. An interesting feature of this approach is that a similar lopsided structure may account for the small mixing in the quark sector\cite{Sato}. Such a possibility, apart from its obvious simplicity, is also well motivated by GUT considerations. In what follows we consider as a framework a class of $SO(10)$ models\cite{Barr,Barr2} with the realistic mass matrices

\begin{equation}
Y^{(u)}\,=\,\left(\begin{array}{ccc}
0 & k' &  0 \\
0 & k  &  b\\
0 & 0  &  a
\end{array}\right)m_u\,,\,\,\,\,\,\,Y^{(N)}=\left(\begin{array}{ccc}
0 & k'&  0 \\
0 & k &  b\\
0 & 0  & a
\end{array}\right)m_u\,,\,
\end{equation}

\begin{equation}
Y^{(d)}\,=\,\left(\begin{array}{ccc}
0 & \delta'- k' & \delta \\
\delta' & -k& \epsilon' -b\\
\delta &\epsilon &  a
\end{array}\right)\,m_d \, ,\,\,\,\,\,\,\,\,Y^{(e)}=\left(\begin{array}{ccc}
0 & \delta'-k' & \delta \\
\delta' & -k&  \epsilon -b\\
\delta &\epsilon' &  a
\end{array}\right)\,m_d\,.
\end{equation}
Only the (common) (33) entry of these matrices, denoted by $a$, is assumed to arise from the standard renormalizable term ${\bf{16}}_3{\bf{16}}_3{\bf{10}}_H$. All other mass entries arise from effective non-renormalizable operators involving additional Higgs fields ${\bf{16}}_H,\,{\bf{16}}_H',\,{\bf{45}}_H$. These contributions are subdominant and are denoted by a number of small parameters ($k,\,k',\,\delta,\,\delta',\,\epsilon,\,\epsilon'$), with the exception of the contribution to the (23) entry, which is assumed to be of the same order as the renormalizable contribution and denoted by the parameter $b$. The small elements $\epsilon,\,\epsilon'$ arise from a non-renormalizable operator $\left\{{\bf{16}}_i{\bf{16}}_H\right\}\left\{{\bf{16}}_j{\bf{16}}_H'\right\}$. The vev $\langle{\bf{16}}'_H\rangle\,\sim\,\langle {N_H^{c}}'\rangle\,\sim M_G$ breaks $SO(10)$ to $SU(5)$, while the vev $\langle{\bf{16}}_H\rangle\,\sim\,\langle H_d^0\rangle\sim\,M_W$ breaks it down to $SU(3)_c\times U(1)_{em}$. Only down quarks and charged leptons get contributions from this term. The relevant Yukawa couplings of this operator respect the $SU(5)$ relation $Y^{(e)}\,=\,\left(Y^{(d)}\right)^{\bot}$, which has been associated with a lopsided structure in the charged lepton sector\cite{Barr}. The  symmetric elements $\delta,\,\delta'$ arise from a different contraction of the same representations, namely $\left\{{\bf{16}}_i{\bf{16}}_j\right\}\left\{{\bf{16}}_H{\bf{16}}_H'\right\}$, appearing again only in $Y^{(d)}$ and $Y^{(e)}$.
A common lopsided structure in the quark and lepton mass matrices arises from the operator $\left\{{\bf{16}}_i{\bf{10}}_H\right\}_{16}\left\{{\bf{16}}_j{\bf{45}}_H\right\}_{\overline{16}}$ through the elements $k,\,k',\,b$. The vev $\langle{\bf{45}}_H\rangle\sim M_G$ lies in the right-handed isospin direction $I_{3R}$, responsible for the breaking of the $SU(2)_R$ subgroup of $SO(10)$, while $\langle{\bf{10}}_H\rangle\,\sim\,M_W$ is the standard vev in the electroweak breaking direction. The contraction employed allows for general Yukawa textures that respect the relation $Y^{(u)}\,=\,Y^{(N)}\,=\,-Y^{(d)}\,=\,-Y^{(e)}$, the minus sign arising from the different $I_{3R}$ charge of the respective fields.

 We proceed by assuming that\cite{Barr} $\delta,\,\delta',\,k,k'\ll\epsilon,\,\epsilon'\ll a,\,b$. Note that out of these parameters one can be absorbed in an overall scale redefinition. Equivalently, here we shall impose the simplifying $b^2+a^2\equiv 1$. Next, by a field redefinition of the down quarks and charged leptons we restrict the complex phases to the (21),(22) and (13) elements, leaving the rest real and positive. Then, without loss of generality, we express (21) and (12) entries in both the down quarks and charged lepton matrices as $Y_{21}\equiv\delta z,\,Y_{12}\equiv|\delta'-k'|$. Furthermore, assuming $\epsilon\sim\epsilon'\ll b$, we approximate the (23) entry as $Y_{23}\approx b$. Thus all parameters besides $z,\delta',k,k'$ are now real in both $Y^{(e)},Y^{(d)}$. Neglecting the overall mass scales, we obtain in this redefined notation (at $M_G$)

\begin{equation}
m_b\,\approx\, m_\tau\,\approx\,(b^2+a^2)^{1/2}\equiv 1,\,\end{equation}
\begin{equation}\,\,m_s/m_b\,\approx\, \epsilon b\,,\,\,\,\,\,m_\mu/m_\tau\,\approx\, \epsilon' b,\,\end{equation}
\begin{equation}|\det{Y^{(d)}}|\,\approx\, |\det{Y^{(e)}}|\,\approx\, |\delta'-k'|\,|z a- b|\,\delta \,.\end{equation}
The model by construction is consistent with the $b-\tau$ unification as a result of the common $b,a$ entries. This is a favourable prediction common in $SO(10)$ and $SU(5)$ models and consistent with the low energy data. To fit the masses $m_s, m_\mu$ of the down quarks and charged leptons we notice that these are controlled by the elements $b\epsilon,\,b\epsilon'$. Then the relation $|\det{Y^{(d)}}|\approx |\det{Y^{(e)}}|$, along with $m_b\approx m_\tau$, results in $m_d/m_e\,\approx\,m_\mu/m_s$ (at $M_G$), which is in general agreement with the expected relevant mass ratios at the unification scale. By taking $\epsilon'/\epsilon\,\approx\,3$, the Georgi-Jarlskog factors can be obtained\cite{Georgi}.

For the up quark masses we have
\begin{equation}
m_c/m_t\,\approx\,\left(|k'|^2+|a k|^2\right)^{1/2}\,,\,\,m_u\,\approx\, 0\,\,.
\end{equation}
The prediction for a massless up quark is, of course, wrong but, since $m_u/m_t\,\sim\,10^{-5}$, a tiny mass for the up quark can always arise from a non-renormalizable operator. Such a small entry in the mass matrices cannot in practice affect the rest of the relations. Furthermore, the parameter $k'(\sim k)$ which appears in both  the mass ratio $m_c/m_t$ and the expression for $\,|\det{Y^{(e)}}|\,$ will allow for a relation between the respective scales.

Next, we notice that since the $M_G$-relation $m_c/m_t\ll m_s/m_b$ is expected to hold, the diagonalization of the up quark matrix will contribute only small corrections to CKM and therefore we can safely consider, in this scheme, quark mixing originating from the down quark matrix. Then, we have the relations
\begin{equation}
V_{cb}\,\approx\,\epsilon a\,,\,\,\,\,V_{ub}\,\approx\, \delta(z^*b+a),\,\end{equation}
\begin{equation} V_{us}\,\approx\,\frac{\delta\epsilon-V_{cb}V_{ub}}{(m_s/m_b)^2},\,\,\,\,\,\,-\frac{V_{ud}V_{ub}^*}{V_{cd}V_{cb}^*}\,\approx\,\frac{b^2(zb+a)}{a-a^2(zb+a)},\,
\end{equation}
where we can easily fit all mixing angles and the CP-violating phase of the quark sector. Using current best-fit values and the expected scale $|\det{Y^{(e)}}|\,\sim \,2\cdot 10^{-5} $, we obtain the rough estimate
\begin{equation}
\left.m_c/m_t\right|_{M_G}\,\approx\,\left(|k'|^2+|a k|^2\right)^{1/2}\sim|\delta'-k'|\,\approx\,4\cdot 10^{-3}\,,\,\,
\end{equation}
within the expected allowed range. An additional important relation is also derived from the quark sector, namely
\begin{equation}
b/a\,\approx\,\frac{m_s/m_b}{|V_{cb}|}\,.
\end{equation}
We are going to see shortly that this ratio will appear as the dominant contribution to $\tan{\theta_{atm}}$ of the neutrino mixing.

 Let us now proceed assuming\footnote{Considering Majorana masses that arise from  $\mathbf{16_i}\,\mathbf{\overline{126}_H}\,\mathbf{16_j}$ or the effective operator $\mathbf{16_i}\,\mathbf{16_j}\,\mathbf{\overline{16}_H}\,\mathbf{\overline{16}_H}$, all Yukawa couplings, without loss of generality, can be expressed in the basis where $Y^{(R)}_{ij}$ is diagonal. } a diagonal Majorana mass matrix for the right-handed neutrinos $Y^{(R)}_\Du\equiv\,diag(1,\,\Lambda,\,1)\,M_R$. The new scale $\Lambda$ is introduced to counteract the large mass hierarchy inherited from the up quark sector to the Dirac neutrino matrix  through the relation $Y^{(u)}=Y^{(N)}$. If this were not the case, the neutrino spectrum would be inconsistent with the observed squared mass differences. The (11) element, taken unity for convenience, is in practice arbitrary as long as the mass ratio $m_{\nu_1}/m_{\nu_3}$ for the light neutrinos, obtained through the seesaw mechanism, is comparable or smaller than $\lambda^4$. We can then manipulate neutrino masses and mixing as previously. Neglecting the overall mass scales, the neutrino matrix defined in (\ref{Y}) is now
\begin{equation}
Y\,\approx\, \left(\begin{array}{ccc}
0\, &\, d' & \,0 \\
\,&\,&\,\\
0\, &\, d\,e^{i\delta_d}\,&\, b\\
\,&\,&\,\\
0\, &\, 0 & \, a
\end{array}\right)\,.
\end{equation}
Since the charged lepton matrix is diagonalized with small rotations, in contrast to the large ones observed in neutrino oscillations, we may consider $U_e\approx I$ to a good approximation. A diagonal phase matrix can then be used to absorb all complex phases (besides the (22) element) and bring the matrix $Y$ into this form. This form is a special case of the {\textit{''Lopsided Neutrino Pattern"}} we previously examined and, thus, using the same treatment we obtain the following relations
\begin{equation}
U_{PMNS}\,\approx\, U_{23}(\theta_{23},0)\,U_{12}(\theta_{12},-\delta_d)\,U'_{23} (\theta'_{23},2\delta_d)\,\cdot {\cal{P}},\,\,\end{equation}
\begin{equation}{\cal{P}}\,=\,diag\left(e^{-i\delta_1},\,e^{-i\delta_d},\,1\right),\,\,\end{equation}
\begin{equation}\tan_{23}\,\approx\, b/a,\,\,\,\tan_{12}\approx d'/(d\cos_{23}),\,\,\tan'_{23}\,\approx\, \frac{d\sin_{23}}{(b^2+a^2)}\left(d'^2+(d\cos_{23})^2\right)^{1/2},\,\end{equation}
\begin{equation}\begin{array}{l}
m_{\nu_3}\,\approx\,| b^2+a^2\,+(d\sin_{23})^2e^{2i\delta_d}|\equiv 1\\
\,\\
m_{\nu_2}/m_{\nu_3}\,\approx\, d'^2+(d\cos_{23})^2\equiv \lambda
\end{array}\,\,\end{equation}
For the diagonal $Y^{(R)}_\Du$ we obtain through the seesaw formula $d'/d = k'/k$. This ratio will allow for a direct fit of the solar angle. We have for the physical parameters
\begin{equation}
\begin{array}{l}
\tan{\theta_{sol}}\,\approx\,\tan_{12}\\
\,\\
\tan{\theta_{atm}}\,\approx\,\left|\tan_{23}+e^{-2i\delta_d}\cos_{12}\tan'_{23}\right|\\
\,\\
|U_{e3}|\,\approx\,\sin_{12}\sin'_{23}
\end{array}{\label{ATM}}
\end{equation}
\begin{equation}
\hat{\delta}_{12}\approx-\delta_1\,,\,\,\,
\hat{\delta}_{13}\approx\delta_d-\delta_1\,,\,\,\,
\hat{\delta}_{23}\approx-\delta_d\,,\,\,\,
\delta_{D}^{\tiny{lep}}\cong 2\delta_d\,.
\end{equation}
From these relations we directly obtain the prediction for the complex phases of the symmetrical parametrization
$\hat{\delta}_{23}+\hat{\delta}_{13}-\hat{\delta}_{12}\approx 0$. By fitting the best-fit value for the ratio $\tan_{23}\approx b/a\approx 0.6$ we notice a significant deviation from the observed atmospheric angle $\tan{\theta_{atm}}\approx 1$, which cannot be accounted for by the subleading term in (\ref{ATM}). An exact fit would require
$m_s/m_b\rightarrow|V_{cb}|$ at $M_G$ and, perhaps, a smaller value  $\theta_{atm}\approx 40^o$ still within current experimental bounds\footnote{Alternatively, by assuming  a large but subleading contribution from  $\left\{{\bf{16}}_3{\bf{10}}_H\right\}_{16}\left\{{\bf{16}}_3{\bf{45}}_H\right\}_{\overline{16}}$  in the (33) entries, the prediction for $b-\tau$ unification is preserved but with the corresponding lepton rotation angle $\tan_{23}=b/a'$ with $a'\sim a$, thus allowing for a direct fit of $\theta_{atm}$.}. Subleading corrections can  also have significant effect on this ratio (especially of our initial working assumption $|\epsilon'-b|\approx b$). In any case, $\delta_{d}$ will be close to zero in this model, giving small CP-violation in neutrino oscillation phenomena but also  $\hat{\delta}_{13}\approx\hat{\delta}_{12}$. The small mixing angle  will obey the relation (\ref{Ue3}) for the corresponding $M_G$ values of the relevant parameters .

\section{Brief Conclusions.}

Summarizing,  we have shown how a lopsided structure hidden within the symmetric light neutrino matrix may account partially or completely for the large lepton mixing angles observed in neutrino oscillation phenomena. Although this idea has been previously considered in other models, here, the assumption of a very light neutrino mass ($m_{\nu_1}/m_{\nu_3}\leq\lambda^4$) allows for an analytic treatment of neutrino masses and mixing. An attractive feature of the formalism developed is that approximations enter only at the stage where the matrix has already been brought to a hierarchical form, thus allowing for exact expressions of the large mixing angles. Among the four instructive lepton patterns  considered, which can potentially fit current lepton mixing data,  the {\textit{``Lopsided neutrino pattern"}}, has a number of appealing features. Specifically, in this model the magnitudes of the lepton mixing angles are predicted within current experimental bounds and the smallness of the $\theta_{13}$ angle is associated with the neutrino mass ratio of the NH case $m_{\nu_2}/m_{\nu_3}\equiv\lambda$. Furthermore, since an analogous lopsided form for the quarks may account for the observed small mixing in the CKM, we also explored the possibility of a common lopsided structure within an $SO(10)$ model with realistic masses and mixing. 
$$$$
{\textbf{Acknowledgements}}

The research presented in this article is co-funded by the European Union - European Social Fund (ESF) $\&$ National Sources, in the framework of the Programme {\textit{"HRAKLEITOS II"}} of the {\textit{"Operational Programme for Education and Lifelong Learning"}} of the {\textit{Hellenic Ministry of Education, Lifelong Learning and Religious Affairs}}. Both authors acknowledge the hospitality of the CERN Theory Group.

\end{document}